\documentclass[]{elsart3p}
\usepackage{graphicx}
\usepackage{amssymb}
\begin{document}
\begin{frontmatter}
\title{Antiferromagnetic, Charge and Orbital Ordered States of Na$_{0.5}$CoO$_{2}$  Based on the Two-Dimensional Triangular Lattice d-p Model}
\author[aff1]{Youichi Yamakawa} and 
\author[aff1,aff2]{Yoshiaki \={O}no}
\address[aff1]{Department of Physics, Niigata University, Ikarashi, Nishi-ku, Niigata 950-2181, Japan}
\address[aff2]{Center for Transdisciplinary Research, Niigata University, Ikarashi, Nishi-ku, Niigata 950-2181, Japan}
%\ead{yamakawa@phys.sc.niigata-u.ac.jp}
%\ead[url]{}
\begin{abstract}

We investigate the electronic state of a CoO$_2$ plane in the layered cobalt oxides Na$_x$CoO$_2$ using the 11 band d-p model on a two-dimensional triangular lattice, where the tight-binding parameters are determined so as to fit the LDA band structure. 
Effects of the Coulomb interaction at a Co site: the intra- and inter-orbital direct terms $U$ and $U'$, the exchange coupling $J$ and the pair-transfer $J'$, are treated within the Hartree-Fock approximation. 
We also consider the effect of the Na order at $x=0.5$, where Na ions form one-dimensional chains, by taking into account of an effective one-dimensional potential $\Delta \varepsilon_{d}$. 
It is found that the one-dimensional Na order enhances the Fermi surface nesting and antiferromagnetism is caused which is suppressed due to the frustration effect in the case without the Na order. 
%It is found that the Na order enhances the Fermi surface nesting resulting in a metallic antiferromagnetism which is suppressed due to the frustration effect in the case without the Na order. 
Furthermore, we consider the effect of the Coulomb interaction between the nearest-neighbor Co sites $V$ and find that a coexistence of the magnetic, charge and orbital ordered state takes place for $V>V_c$ where the system becomes insulator. 
\end{abstract}
\end{frontmatter}
%%%%%%%%%%%%%%%%%%%%%%%%%%%%%%%%%%%%%%%%%%%%%%%%%%%%%%%%%%%%%%%%%%%%%%%%%%%%
\section{Introduction}
The discovery of Superconductivity in the layered cobalt oxide Na$_{x}$CO$_{2} \cdot$yH$_{2}$O \cite{Takada} has stimulated further interest in the electronic states of the mother compound Na$_x$CoO$_2$. 
The specific features of the system are the geometrical fluctuation of the CoO$_2$ plane which consists of a triangular lattice of Co atoms, and the orbital degeneracy of $t_{2g}$ bands of Co $3d$ electrons. 
The electron filling in the CoO$_2$ plane is controlled by changing the Na content $x$: the hole concentration of Co $t_{2g}$ bands is given by $n_{\rm hole}=1-x$. 

For $x \leq 0.6$, Na$_x$CoO$_2$ is a normal paramagnetic metal, while, for $0.6 \leq x \leq 0.75$, an anomalous behavior is observed: the magnetic susceptibility $\chi$ is Curie-Weiss like although the resistivity is metallic \cite{Foo}, the electronic specific heat coefficient $\gamma$ is large and increases with $x$ \cite{Yokoi}, the thermopower is unusually large \cite{Terasaki}. 
For $x \geq 0.75$, a weak magnetic order is observed below 20K \cite{Motohashi}, where the ferromagnetic ordered CoO$_2$ layers couple antiferromagnetically with each other \cite{Bayrakci}. 
%When H$_2$O is intercalated between the CoO$_2$ layers, Takada {\it et. al.} \cite{Takada} have discovered the superconductivity in Na$_{x}$CoO$_2\cdot y$H$_2$O with $T_c \sim 5$ K for $x \approx 0.35$ and $y \approx 1.3$. 

Furthermore, Na$_{0.5}$CoO$_2$ exhibits a remarkable successive phase transition at $T_{c1}\sim 87$ K and  $T_{c2}\sim 53$ K \cite{Foo,Yokoi}: the in-plane antiferromagnetic order is realized below $T_{c1}$ and the system becomes insulator below $T_{c2}$. 
From the NMR and the Neutron measurements, Yokoi {\it et. al.} \cite{Yokoi} have proposed the magnetic structure of Na$_{0.5}$CoO$_2$, where chains of Co$^{3.5+\delta}$ with larger staggered moment and Co$^{3.5-\delta}$ with smaller moment exist alternatively within the CoO$_2$ plane. 
The charge ordering of the Co sites into chains of Co$^{3.5+\delta}$ and Co$^{3.5-\delta}$ is closely related to the ordered pattern of Na ions which form one-dimensional chains below room temperature \cite{Huang}. 
In spite of the intense effort devoted in the last few years, the origin of metal-insulator transition at $T_{c2}$ is still an open question. 

The purpose of this paper is to investigate the electronic state of the CoO$_2$ plane including the successive phase transition of Na$_{0.5}$CoO$_2$, particularly focused on the effect of the one-dimensional Na order on the antiferromagnetism. 
For this purpose, we take into account of an effective one-dimensional potential on the CoO$_2$ plane due to the effect of the Na order. 

The contents of this paper are as follows; 
In Sec. 2, we describe the model Hamiltonian for the CoO$_{2}$ plane. 
In Sec. 3, results of the calculation are presented. 
Finally, we summarize the results of the present work in Sec. 4. 
%%%%%%%%%%%%%%%%%%%%%%%%%%%%%%%%%%%%%%%%%%%%%%%%%%%%%%%%%%%%%%%%%%%%%%%%%%%%
\section{Model}
To investigate the electronic states of the CoO$_2$ plane in the layered cobalt oxides Na$_{x}$CoO$_2$, we employ the two dimensional triangular lattice $d$-$p$ model which includes 11 orbitals:  $d_{xy}$, $d_{yz}$, $d_{zx}$, $d_{x^2-y^2}$, $d_{3z^2-r^2}$ of Co and $p_{1x}$, $p_{1y}$, $p_{1z}$ ($p_{2x}$, $p_{2y}$, $p_{2z}$) of O in the upper (lower) side of a Co plane.
The Hamiltonian is given by; 
%%%%%%%%%%%%%%%%%%%%%%%%%%%%%%%%%%%%%%%%%%%%%%%%%%%%%%%%%%%%%%%%%%%%%%%%%%%%
\begin{eqnarray}
H      &=& H_{p} + H_{dp} + H_{d} + H_{U} + H_{V} \,, \label{eq-Model} \\
H_{p}  &=& \varepsilon_{p} \sum_{{\bf k},j,l,\sigma}
           p_{{\bf k},j,l,\sigma}^{\dagger}
           p_{{\bf k},j,l,\sigma} \nonumber \\
       &&+ \!\!\!\! \sum_{{\bf k},j,j',l,l',\sigma}
           \!\!\!\! t_{{\bf k},j,j',l,l'}^{pp}
           p_{{\bf k},j,l,\sigma}^{\dagger}
           p_{{\bf k},j',l',\sigma} \,,  \label{eq-Model2} \\
H_{dp} &=& \sum_{{\bf k},j,l,m}
          (t_{{\bf k},j,l,m}^{pd}
           p_{{\bf k},j,l,\sigma}^{\dagger}
           d_{{\bf k},m,\sigma}
         + {\rm h.c.}) \,,  \label{eq-Model3} \\
H_{d}  &=& \sum_{{\bf n},m,\sigma}
           \varepsilon_{{\bf n},m}^{d}
           d_{{\bf n},m,\sigma}^{\dagger}
           d_{{\bf n},m,\sigma} \nonumber \\
       &&+ \sum_{{\bf k},m,m',\sigma}
           t_{{\bf k},m,m'}^{dd}
           d_{{\bf k},m,\sigma}^{\dagger}
           d_{{\bf k},m',\sigma}  \label{eq-Model4} \\
H_{U}  &=& U \sum_{{\bf n},m}
           d_{{\bf n},m,\uparrow}^{\dagger}
           d_{{\bf n},m,\uparrow}
           d_{{\bf n},m,\downarrow}^{\dagger}
           d_{{\bf n},m,\downarrow} \nonumber \\
       &&+ U'\sum_{{\bf n},m \neq m'}
           d_{{\bf n},m,\uparrow}^{\dagger}
           d_{{\bf n},m,\uparrow}
           d_{{\bf n},m',\downarrow}^{\dagger}
           d_{{\bf n},m',\downarrow} \nonumber \\
       &&+ \left( U' - J \right) \!\!\!\! 
           \sum_{{\bf n},m>m',\sigma} \!\!\!\! 
           d_{{\bf n},m,\sigma}^{\dagger}
           d_{{\bf n},m,\sigma}
           d_{{\bf n},m',\sigma}^{\dagger}
           d_{{\bf n},m',\sigma} \nonumber \\
       &&+ J \sum_{{\bf n},m \neq m'}
           d_{{\bf n},m,\uparrow}^{\dagger}
           d_{{\bf n},m',\uparrow}
           d_{{\bf n},m',\downarrow}^{\dagger}
           d_{{\bf n},m,\downarrow} \nonumber \\
       &&+ J' \sum_{{\bf n},m \neq m'}
           d_{{\bf n},m,\uparrow}^{\dagger}
           d_{{\bf n},m',\uparrow}
           d_{{\bf n},m,\downarrow}^{\dagger}
           d_{{\bf n},m',\downarrow} \,,  \label{eq-Model5} \\
H_{V}  &=& V \!\!\!\! \!\!\!\!  \sum_{\left< {\bf n,n'} \right>
                   ,m,m',\sigma,\sigma '}\!\!\!\! \!\!\!\! 
           d_{{\bf n}, m,\sigma}^{\dagger}
           d_{{\bf n}, m,\sigma}
           d_{{\bf n'}, m',\sigma '}^{\dagger}
           d_{{\bf n'}, m',\sigma '} \,,  \label{eq-Model6}
\end{eqnarray}
%%%%%%%%%%%%%%%%%%%%%%%%%%%%%%%%%%%%%%%%%%%%%%%%%%%%%%%%%%%%%%%%%%%%%%%%%%%%
where $d_{{\bf k},m,\sigma}^{\dagger}$ ($d_{{\bf n},m,\sigma}^{\dagger}$) is a creation operator for a cobalt $3d$ electron with wave vector ${\bf k}$ (site ${\bf n}=(n_x,n_y)$), orbital $m$ ($xy$, $yz$, $zx$, $x^2-y^2$, $3z^2-r^2$) and spin $\sigma$,  and $p_{{\bf k},j,l,\sigma}^{\dagger}$ is a creation operator for a oxygen $2p$ electron with  wave vector ${\bf k}$, site $j(=1,2)$, orbital $l$ ($x$, $y$, $z$) and spin $\sigma$, respectively. 
In Eqs. (\ref{eq-Model2}), (\ref{eq-Model3}) and (\ref{eq-Model4}), the transfer integrals $t_{{\bf k},j,j',l,l'}^{pp}$, $t_{{\bf k},j,l,m}^{pd}$ and $t_{{\bf k},m,m'}^{dd}$, which are written by the Slater-Koster parameters, together with the atomic energies $\varepsilon_{p}$ and $\varepsilon_{m}^{d}$ are determined so as to fit the tight-binding energy bands to the LDA bands for Na$_{0.5}$CoO$_2$ \cite{Singh}, and to be suitable for the result of ARPES \cite{Yang} parameters are slightly adjusted. 
In Eq. (\ref{eq-Model6}), the sum run over the nearest neighbor pairs $\left< {\bf n,n'} \right>$ over all lattice sites.

In the Hamiltonian Eq. (\ref{eq-Model}), we consider the effects of the intra-atomic Coulomb interaction at a Co site: the intra- and inter-orbital direct terms $U$ and $U'$, the exchange coupling $J$ and the pair-transfer $J'$ \cite{Mizokawa}, and effects of the Inter-atomic Coulomb interaction between the nearest neighbor Co sites $V$. 
Here and hereafter, we assume the rotational invariance yielding the relations: $U'=U-2J$ and $J=J'$. 
The effect of the one-dimensional Na order, in which Na ions form one-dimensional chains at $x=0.5$, is considered by taking into account an effective one-dimensional potential on the CoO$_2$ plane: 
%%%%%%%%%%%%%%%%%%%%%%%%%%%%%%%%%%%%%%%%%%%%%%%%%%%%%%%%%%%%%%%%%%%%%%%%%%%%
\begin{eqnarray}
\label{eq-Na}
   \varepsilon_{{\bf n},m}^{d}
 = \left\{
   \begin{array}{ll}
      \varepsilon_{m}^{d} - \Delta \varepsilon_{d}
    & \begin{array}{l}
         \qquad $for odd $n_y \\
         $(on the Na ordered line) $ \\
      \end{array} \\
      \varepsilon_{m}^{d} + \Delta \varepsilon_{d}
    & \begin{array}{l}
         \qquad $for even $n_y\\
         $(out of the Na ordered line),$ \\
      \end{array}
   \end{array}
   \right.
\end{eqnarray}
%%%%%%%%%%%%%%%%%%%%%%%%%%%%%%%%%%%%%%%%%%%%%%%%%%%%%%%%%%%%%%%%%%%%%%%%%%%%
with the effective potential $\Delta \varepsilon_{d}$ due to the Na order. 
Because a part of the effect of the inter-atomic Coulomb interaction $V$ enhances $\Delta \varepsilon_{d}$, $V$ is corrected so as not to influence $\Delta \varepsilon_{d}$. 
Hereafter, in order to consider the both Na order and the antiferromagnetic order, we use an extended unit cell including four Co atoms and eight O atoms, together with the magnetic Brillouin zone which is $\frac14$ of the original Brillouin zone. 

%%%%%%%%%%%%%%%%%%%%%%%%%%%%%%%%%%%%%%%%%%%%%%%%%%%%%%%%%%%%%%%%%%%%%%%%%%%%
\section{Result}
Now we discuss possible ordered states including magnetic, charge and orbital  ordered states in the case of $x=0.5$ within the Hartree-Fock approximation \cite{Mizokawa}. 
In this study, we assume that the order parameters are diagonal with respect to the orbital $m$ and the spin $\sigma$. 

\begin{figure}[h]
\begin{center}
\includegraphics[height=3.80cm]{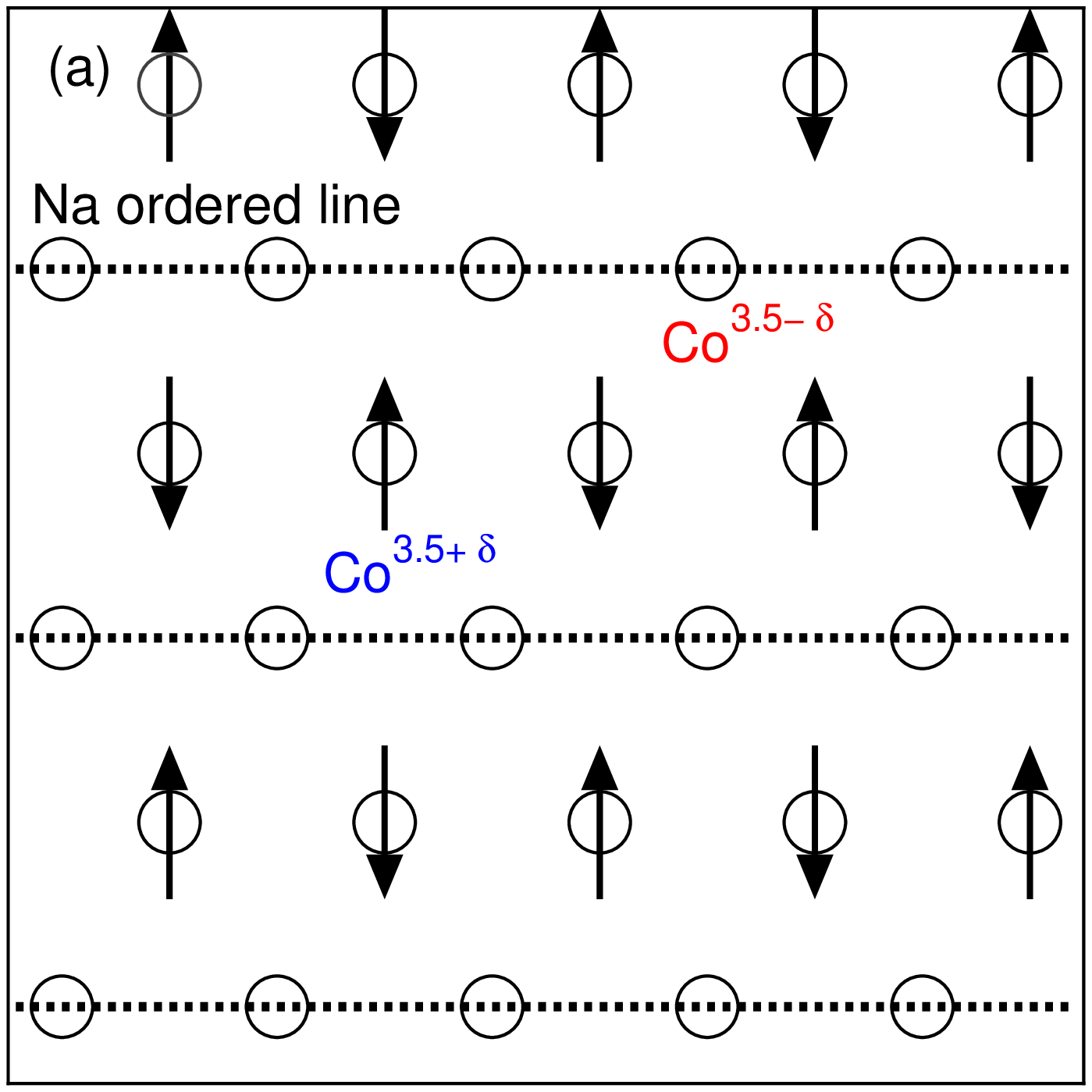}
\includegraphics[height=3.80cm]{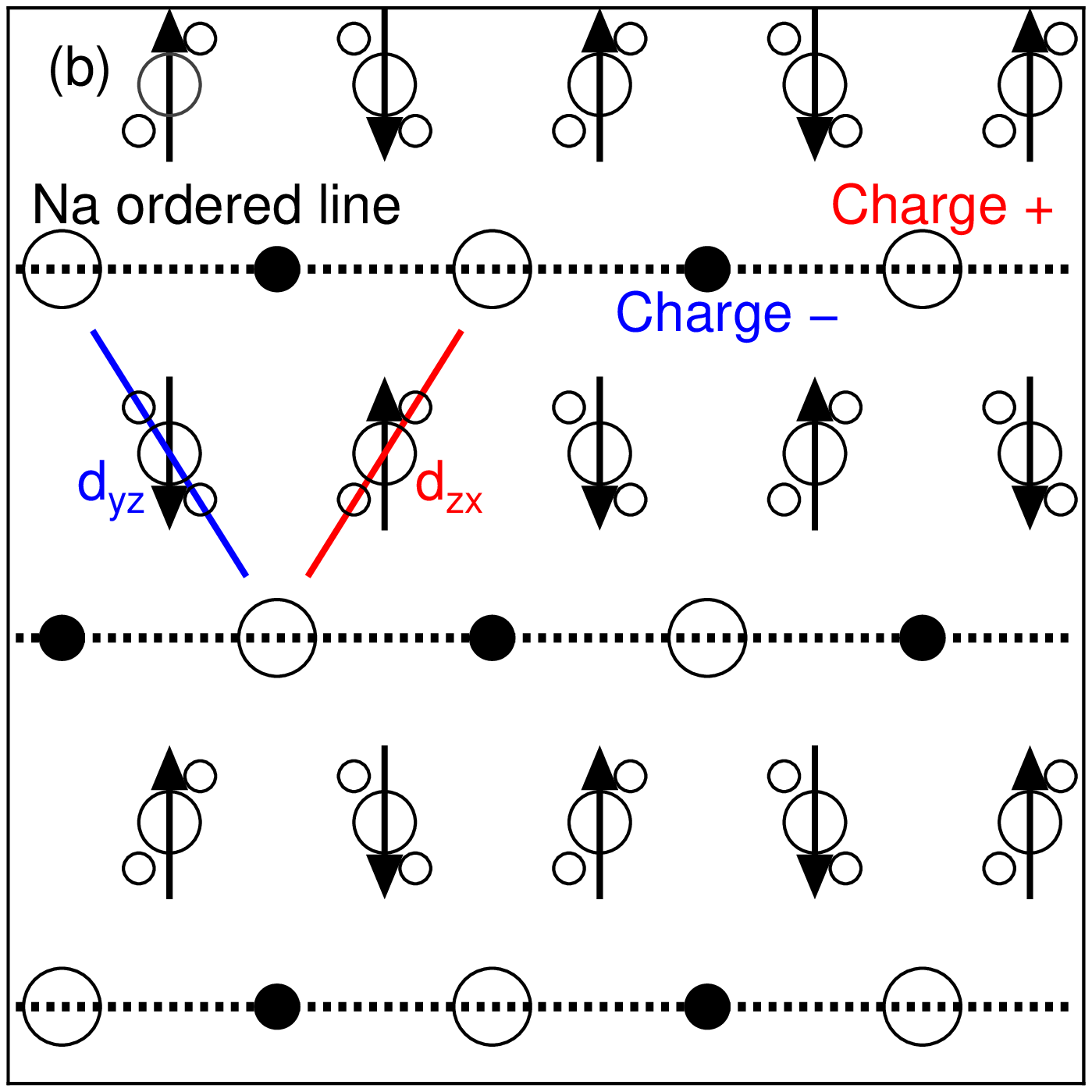}
\caption{
Order patterns. The left panel is the antiferromagnetic ordered state (a), and the right panel is the coexistence state between the antiferromagnetic, charge and orbital ordered state (b). 
}
\label{fig-struct}
\end{center}
\end{figure}

Figure \ref{fig-struct}(a) shows a possible order pattern of the antiferromagnetic state.  Chains of Co 3.5+$\delta$ and Co 3.5-$\delta$ exist alternatingly within the CoO$_2$ plane because of the one-dimensional Na order \cite{Huang}. 
Figure \ref{fig-struct}(b) shows an order pattern of the coexistence state of antiferromagnetic, charge and orbital order, where
the Co sites out of the Na ordered line with the valency of 3.5+$\delta$ have the antiferromagnetic moment \cite{Yokoi}, while, the Co sites on the Na ordered line with the valency of 3.5-$\delta$ show the charge order which is accompanied by the orbital order out of the Na ordered line. 
The coexistence state of the antiferromagnetic and charge order without the orbital order has not been obtained in the present calculation. 

%%%%%%%%%%%%%%%%%%%%%%%%%%%%%%%%%%%%%%%%%%%%%%%%%%%%%%%%%%%%%%%%%%%%%%%%%%%%
\vspace{0.5cm}
%%%%%%%%%%%%%%%%%%%%%%%%%%%%%%%%%%%%%%%%%%%%%%%%%%%%%%%%%%%%%%%%%%%%%%%%%%%%

\begin{figure}[h]
\begin{center}
\includegraphics[height=4.0cm]{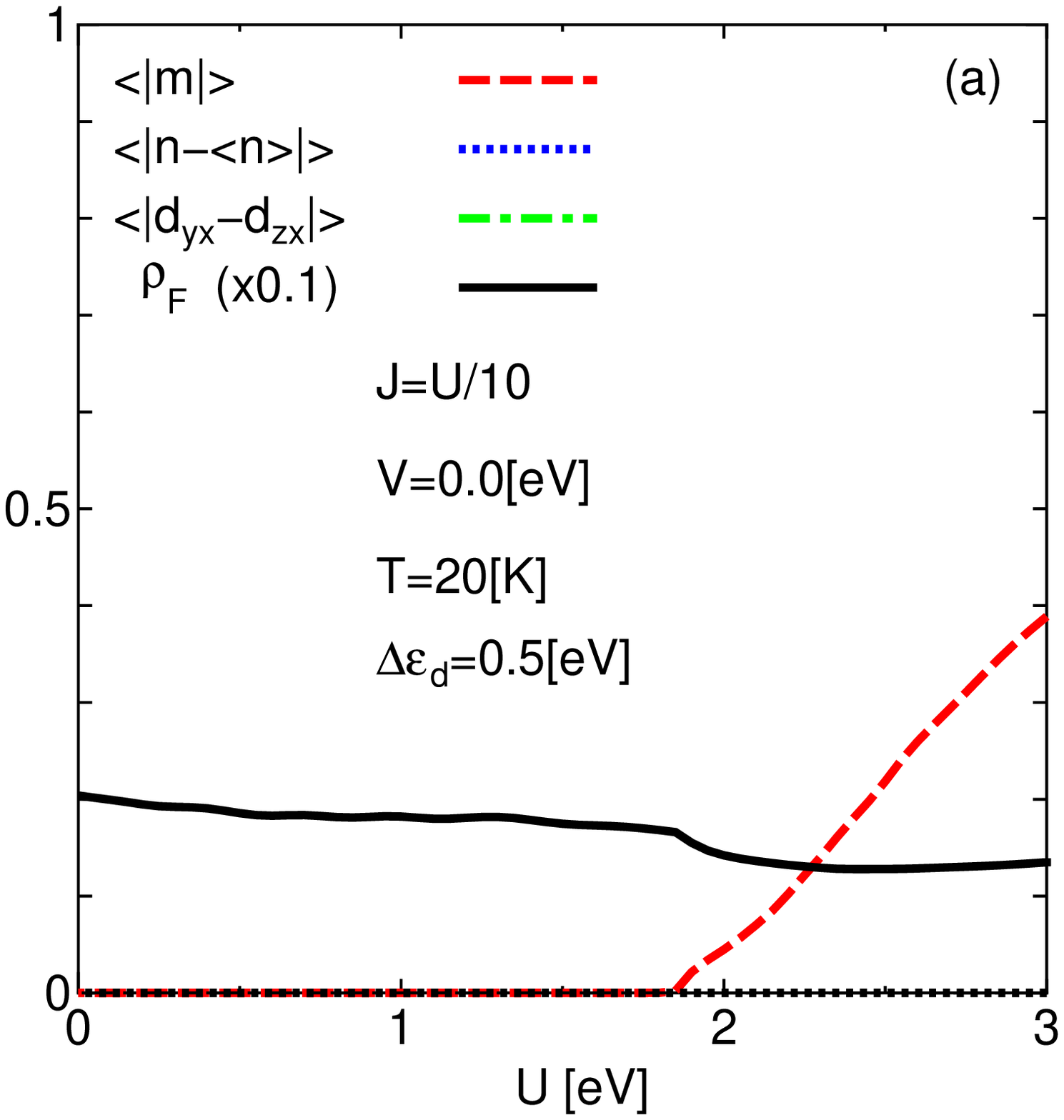}
\includegraphics[height=4.0cm]{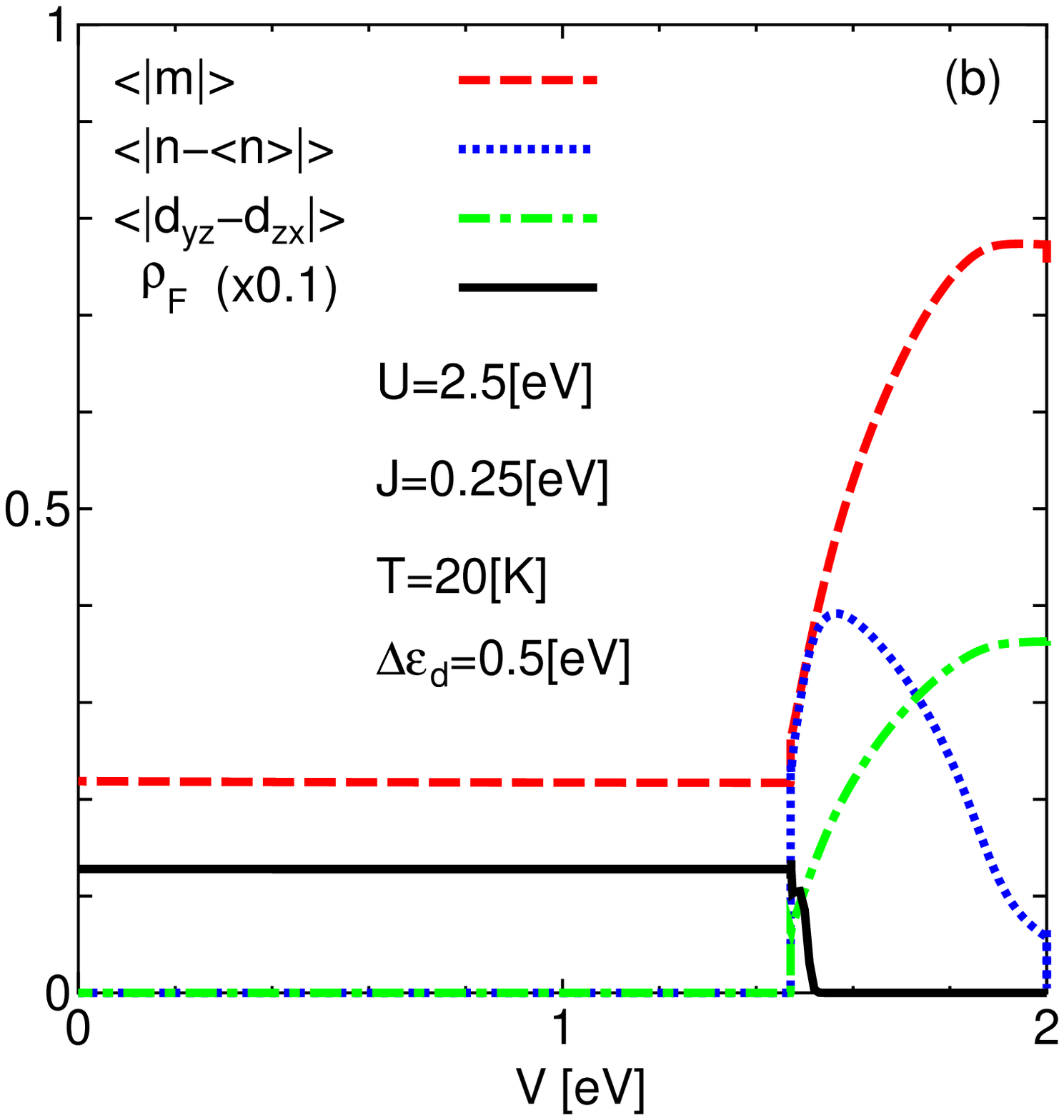}
\caption{
Order parameters. Sublattice magnetization $m$, charge order $n-\left< n \right>$, orbital order $n_{yz}-n_{zx}$ and the density of states $\rho_{\rm F}$ are shown as functions of the intra-atomic Coulomb interaction $U$ (a) and inter-atomic Coulomb interaction $V$ (b). 
}
\label{fig-phase}
\end{center}
\end{figure}

\begin{figure}[h]
\begin{center}
\includegraphics[height=3.5cm,angle=-90]{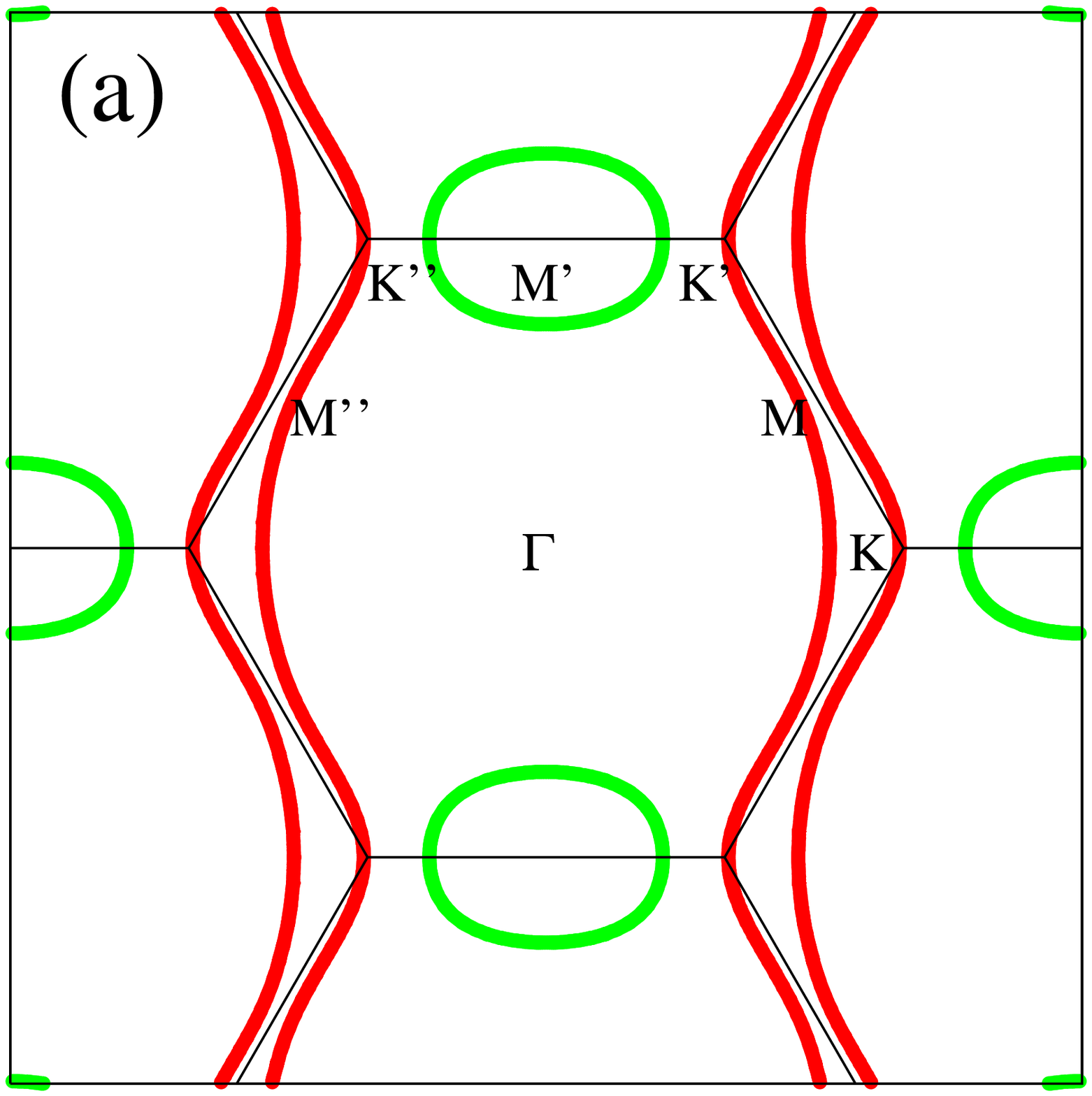}
\includegraphics[height=4.2cm,angle=-90]{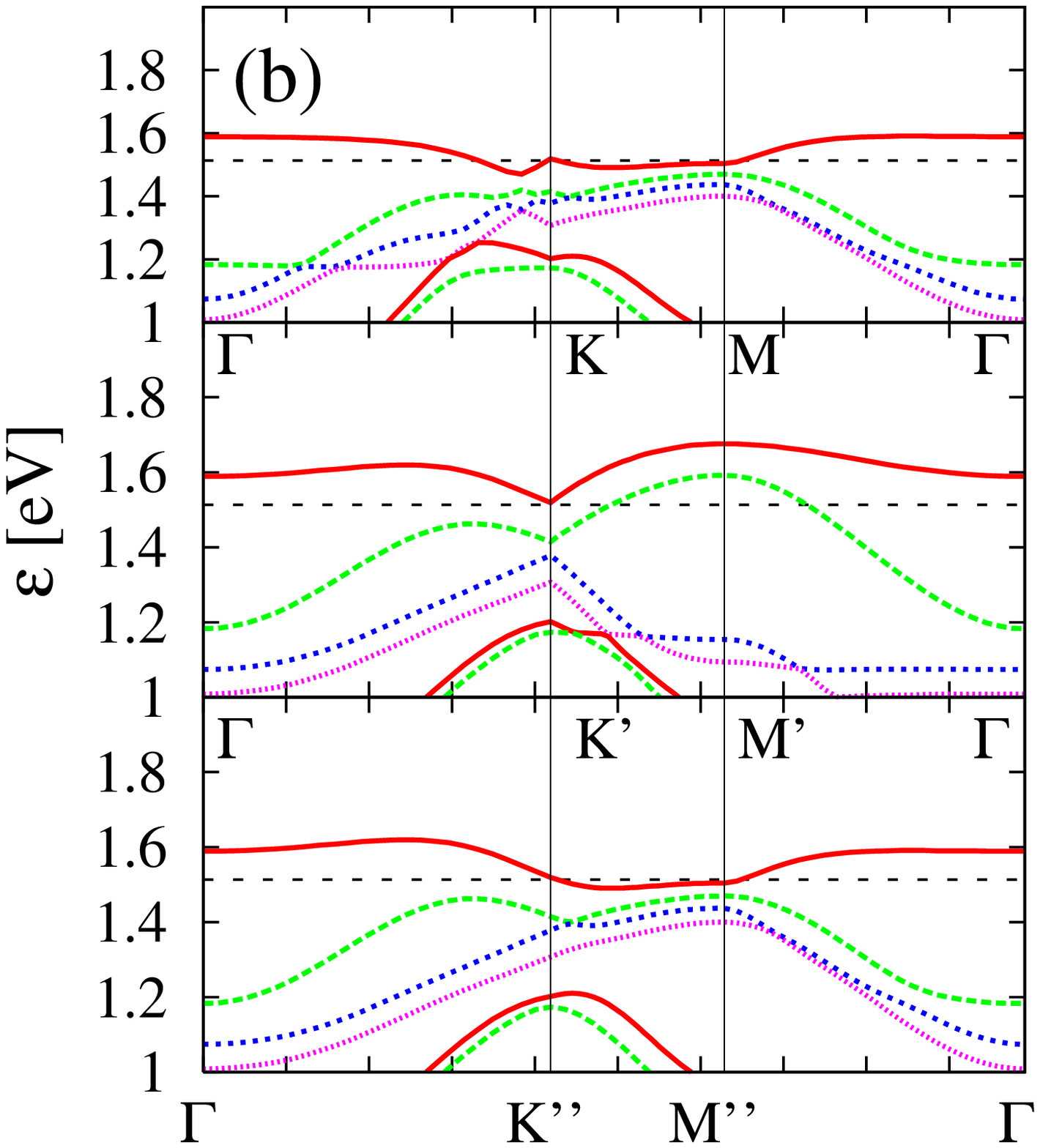}
\caption{
Fermi surfaces (a) and band structure (b) in the magnetic Brillouin zone for the normal state at $U=J=V=0$ eV, $T=20$ K and $\Delta \varepsilon_{d} = 0.5$ eV. 
}
\label{fig-fermi0}
\end{center}
\end{figure}

Figure \ref{fig-phase}(a) shows the sublattice magnetization $m$, charge order $n-\left< n \right>$, orbital order $n_{yz}-n_{zx}$ and the density of states at Fermi level $\rho_{\rm F}$ as functions of $U$ with $V=0$ eV, temperature $T=20$ K and the effective potential $\Delta \varepsilon_{d}=0.5$ eV. 
The phase transition of antiferromagnetic ordered state occurs at $U_{c} \sim 1.8$ eV. 
With increasing $U$, sublattice magnetization $m$ increases, while, the density of states $\rho_{\rm F}$ is almost constant. 
Due to the effect of the one-dimensional potential, the band structure becomes quasi one-dimensional and the Fermi surface nesting is enhanced as shown in Fig. \ref{fig-fermi0}. 
Then, the antiferromagnetic ordered state appears in the presence of the Na order \cite{Yamakawa}. 

Figure \ref{fig-phase}(b) shows $m$, $n-\left< n \right>$, $n_{yz}-n_{zx}$ and $\rho_{\rm F}$ as functions of $V$ with $U=2.5$ eV, $J=0.25$ eV, $T=20$ K and $\Delta \varepsilon_{d}=0.5$ eV. 
The phase transition from the metallic antiferromagnetic ordered state to the insulating coexistence state is obtained at $V_{c} \sim 1.5$ eV. 
At $V_{c}$, the charge order at Co sites on the Na ordered line is caused by the effect of $V$, and the anisotropy between $d_{yz}$ and $d_{zx}$ at Co sites out of the Na ordered line is caused by the charge order as shown in Fig. \ref{fig-struct}. 

%%%%%%%%%%%%%%%%%%%%%%%%%%%%%%%%%%%%%%%%%%%%%%%%%%%%%%%%%%%%%%%%%%%%%%%%%%%%
%%%%%%%%%%%%%%%%%%%%%%%%%%%%%%%%%%%%%%%%%%%%%%%%%%%%%%%%%%%%%%%%%%%%%%%%%%%%

In Fig. \ref{fig-fermi0}, the Fermi surface and the renormalized band structure of the normal state  are shown at $U=J=V=0$, $T=20$ K and $\Delta \varepsilon_{d} = 0.5$ eV. 
The effect of the one-dimensional potential $\Delta \varepsilon_{d}$ due to the Na order enhances the Fermi surface nesting. 

\begin{figure}[h]
\begin{center}
\includegraphics[height=3.5cm,angle=-90]{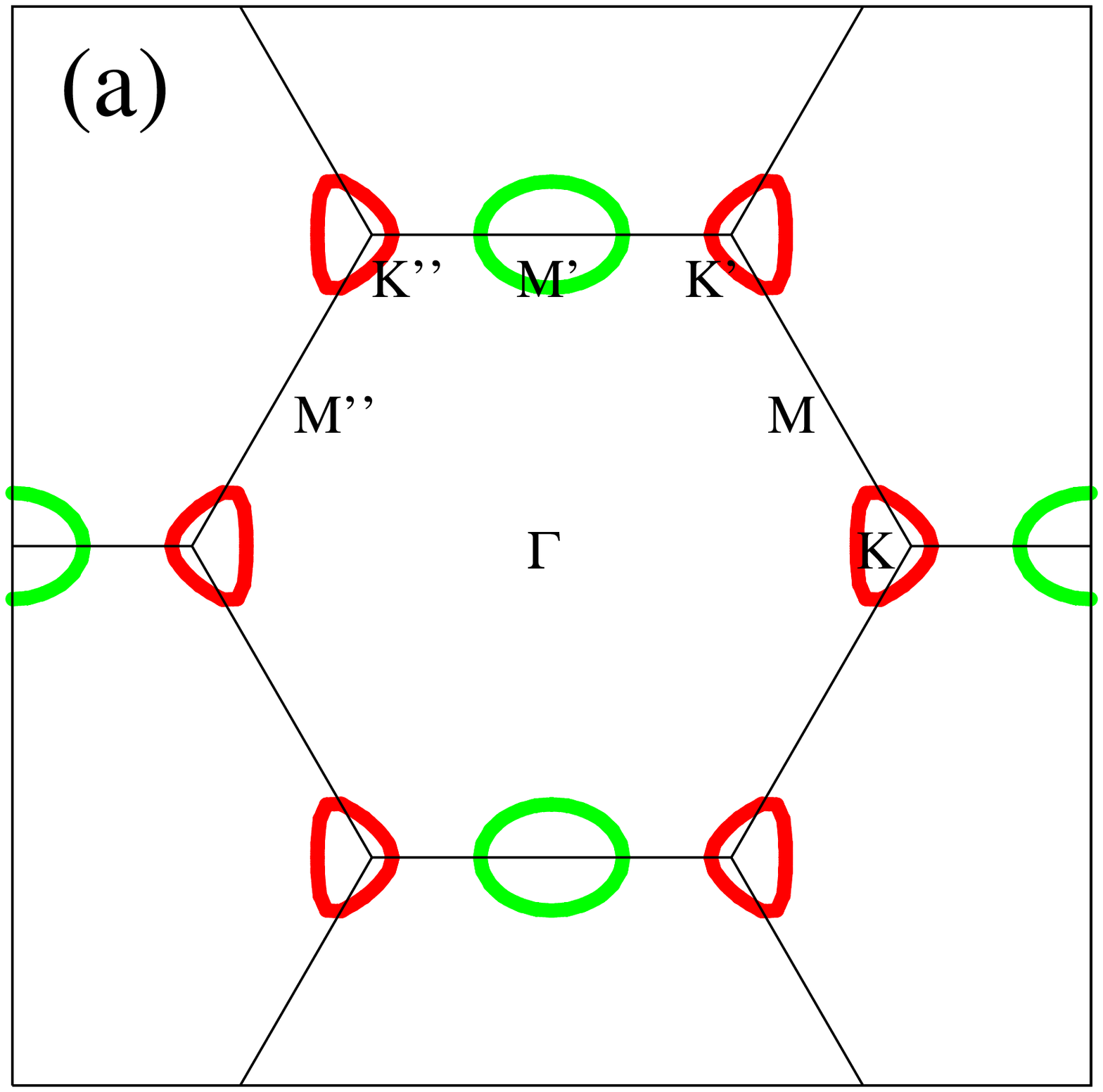}
\includegraphics[height=4.2cm,angle=-90]{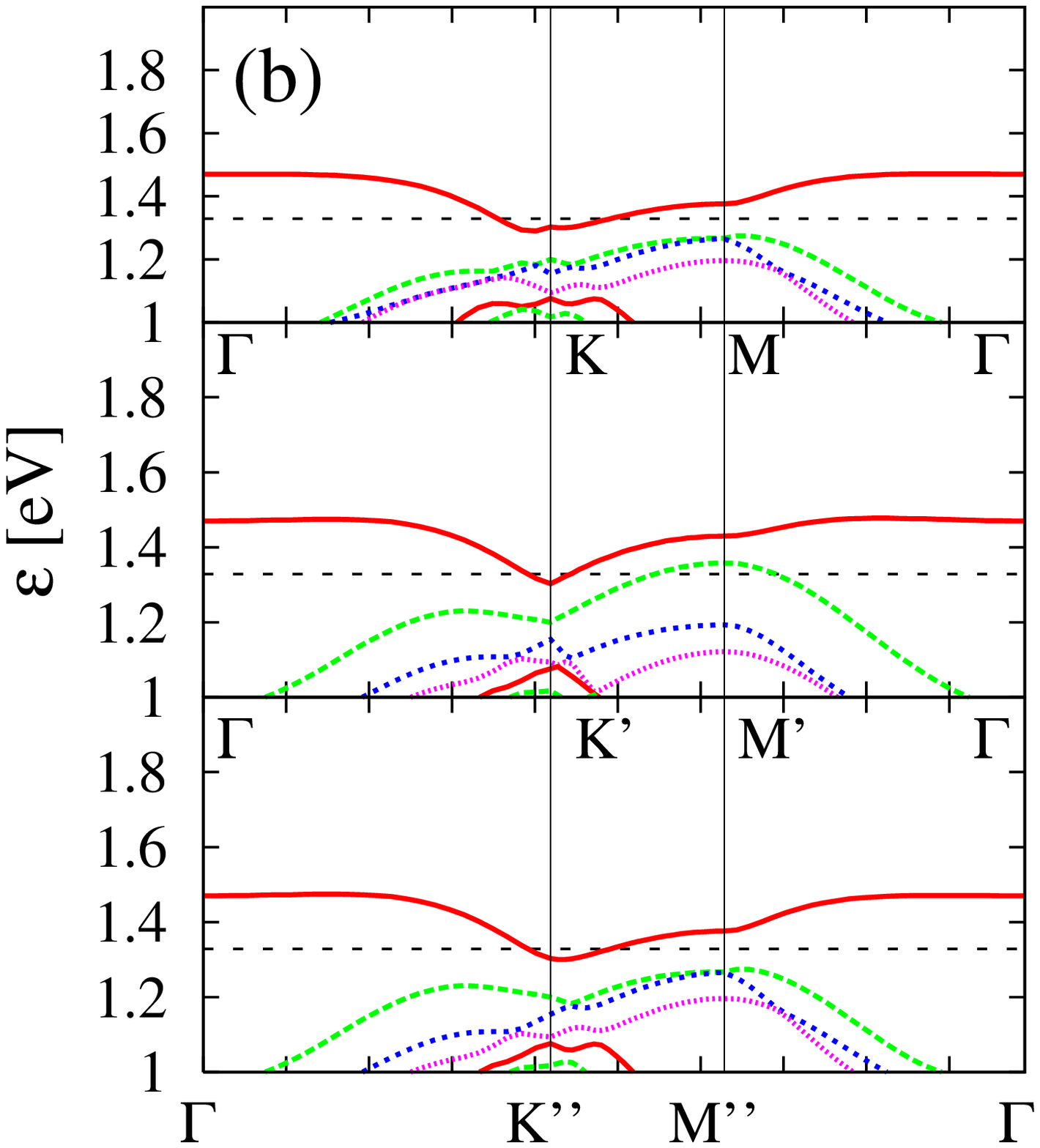}
\caption{
Fermi surfaces (a) and band structure (b) for the metallic antiferromagnetic state at $U=2.5$ eV, $J=0.25$ eV, $V=0$ eV, $T=20$ K and $\Delta \varepsilon_{d} = 0.5$ eV. 
}
\label{fig-fermi1}
\end{center}
\end{figure}

In Fig. \ref{fig-fermi1}, we plot the Fermi surface and the renormalized band structure in the antiferromagnetic ordered state at $U=2.5$ eV, $J=0.25$ eV, $V=0$ eV, $T=20$ K and $\Delta \varepsilon_{d} = 0.5$ eV. 
This state is metallic because the electron like Fermi surfaces around K (K', K'') points and the hole like Fermi surfaces around M' point yield a large value of the density of states $\rho_{\rm F}$. 

\begin{figure}[h]
\begin{center}
\includegraphics[height=3.5cm,angle=-90]{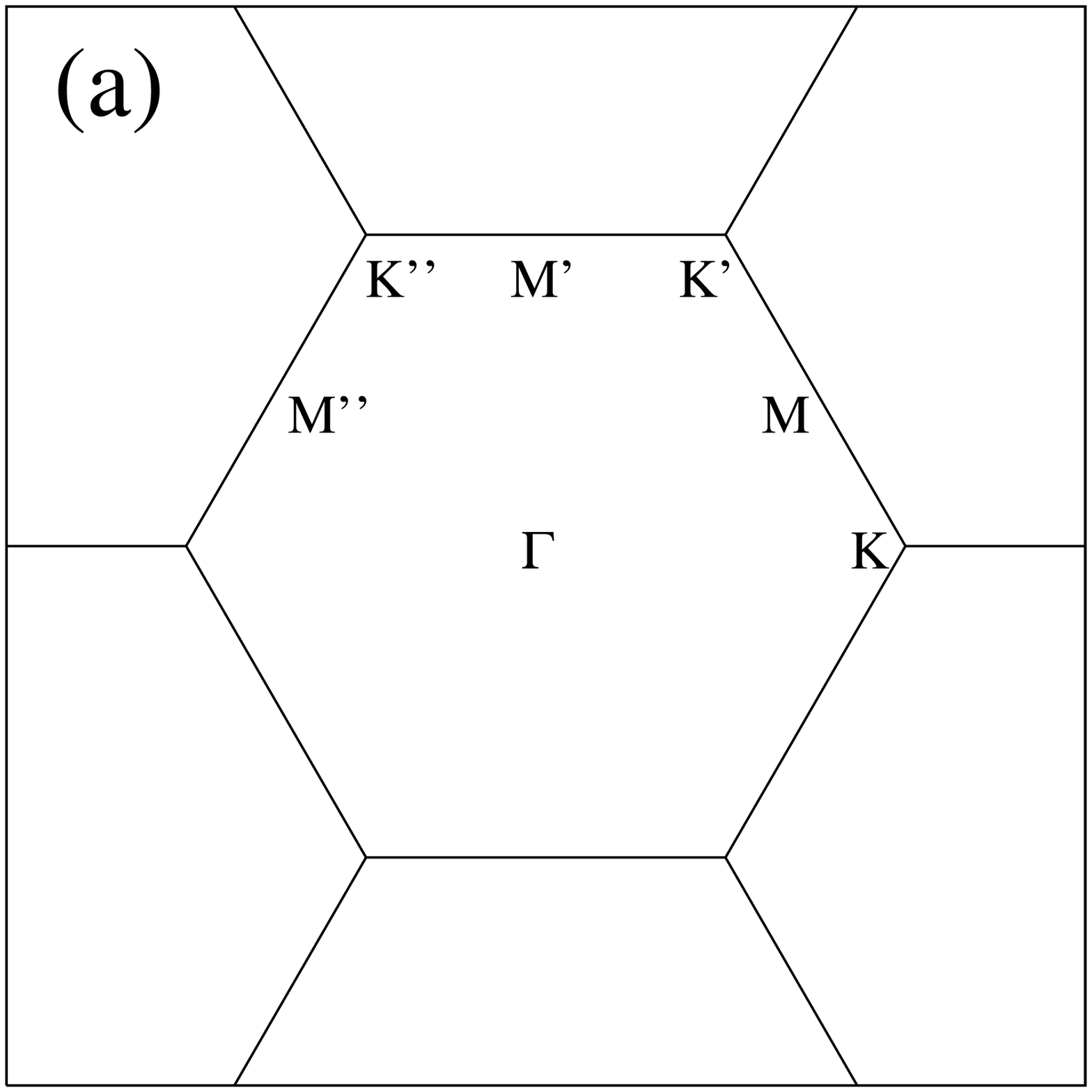}
\includegraphics[height=4.2cm,angle=-90]{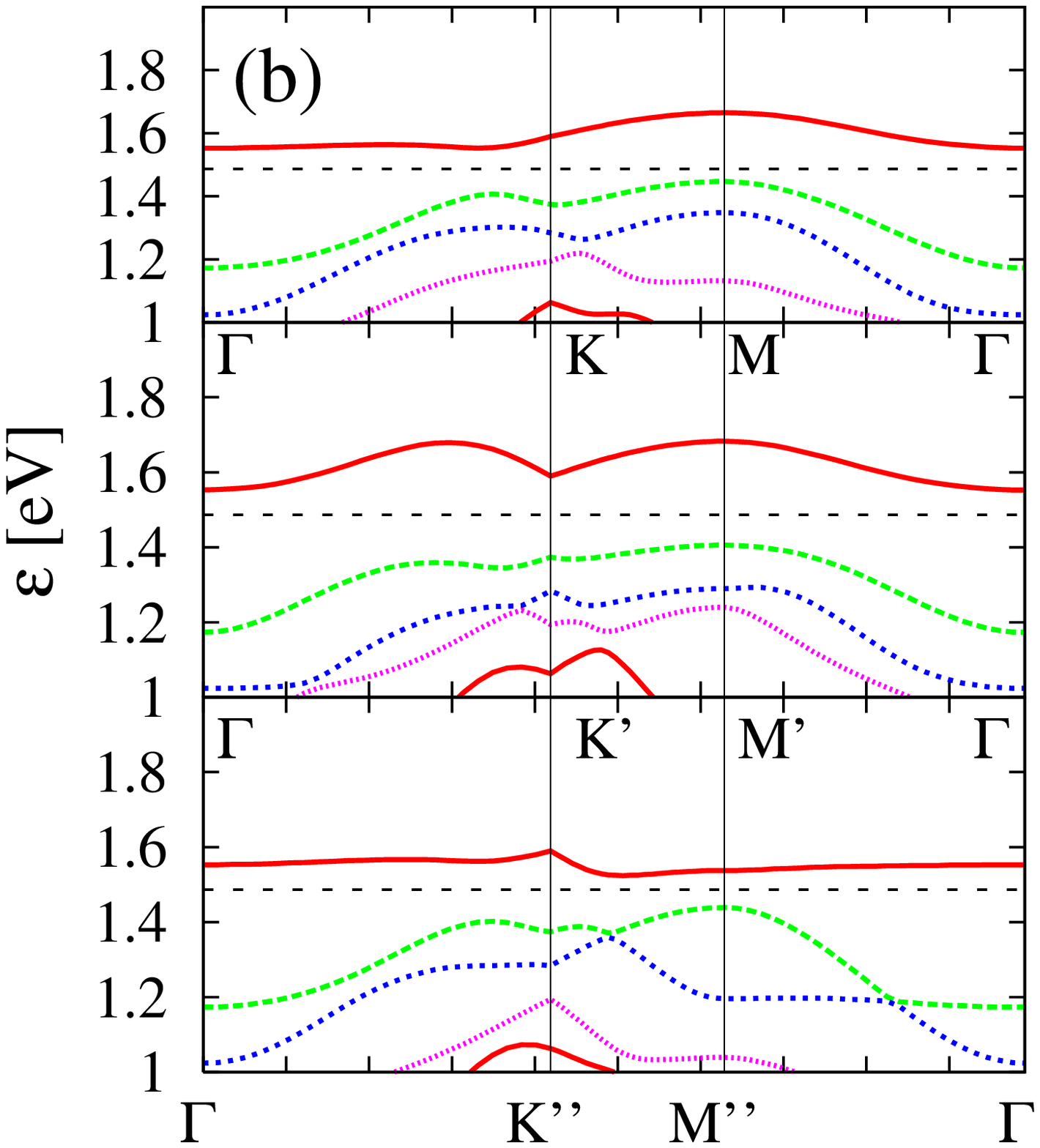}
\caption{
Fermi surfaces (a) and band structure (b) for the insulating coexistence state with finite energy-gap at $U=2.5$ eV, $J=0.25$ eV, $V=1.60$ eV, $T=20$ K and $\Delta \varepsilon_{d} = 0.5$ eV. 
}
\label{fig-fermi3}
\end{center}
\end{figure}

The Fermi surface and the renormalized band structure of the coexistence state are shown in Fig. \ref{fig-fermi3}. 
This state is insulating because the renormalized band has a finite energy gap. 

%%%%%%%%%%%%%%%%%%%%%%%%%%%%%%%%%%%%%%%%%%%%%%%%%%%%%%%%%%%%%%%%%%%%%%%%%%%%
%%%%%%%%%%%%%%%%%%%%%%%%%%%%%%%%%%%%%%%%%%%%%%%%%%%%%%%%%%%%%%%%%%%%%%%%%%%%
%%%%%%%%%%%%%%%%%%%%%%%%%%%%%%%%%%%%%%%%%%%%%%%%%%%%%%%%%%%%%%%%%%%%%%%%%%%%
%%%%%%%%%%%%%%%%%%%%%%%%%%%%%%%%%%%%%%%%%%%%%%%%%%%%%%%%%%%%%%%%%%%%%%%%%%%%
%%%%%%%%%%%%%%%%%%%%%%%%%%%%%%%%%%%%%%%%%%%%%%%%%%%%%%%%%%%%%%%%%%%%%%%%%%%%
% \begin{figure}[h]
% \begin{center}
% \includegraphics[height=3.5cm,angle=-90]{ps-fermi-a005b052.ps}
% \includegraphics[height=4.2cm,angle=-90]{ps-band-a005b052.ps}
% \caption{
% Fermi surfaces (a) and band structure (b) in the magnetic Brillouin zone for the coexistence state at $U=2.5$ eV, $J=0.25$ eV, $V=1.48$ eV, $T=20$ K and $\Delta \varepsilon_{d} = 0.5$ eV. 
% }
% \label{fig-fermi2}
% \end{center}
% \end{figure}
%%%%%%%%%%%%%%%%%%%%%%%%%%%%%%%%%%%%%%%%%%%%%%%%%%%%%%%%%%%%%%%%%%%%%%%%%%%%
%%%%%%%%%%%%%%%%%%%%%%%%%%%%%%%%%%%%%%%%%%%%%%%%%%%%%%%%%%%%%%%%%%%%%%%%%%%%
%%%%%%%%%%%%%%%%%%%%%%%%%%%%%%%%%%%%%%%%%%%%%%%%%%%%%%%%%%%%%%%%%%%%%%%%%%%%

%%%%%%%%%%%%%%%%%%%%%%%%%%%%%%%%%%%%%%%%%%%%%%%%%%%%%%%%%%%%%%%%%%%%%%%%%%%%
\section{Summary and Discussion}
We investigated the electronic state in the CoO$_2$ plane of the layered cobalt oxides Na$_{0.5}$CoO$_2$ by using the 11 band $d$-$p$ model on the two-dimensional triangular lattice within the Hartree-Fock approximation. 
We obtained two ordered states: 
(1) The metallic antiferromagnetic ordered state for $U>U_{c}$. 
(2) The insulating coexistence state between antiferromagnetic, charge and orbital ordered state for $U>U_{c}$ and $V>V_{c}$. 
It is found that the one-dimensional Na order enhances the Fermi surface nesting around M and M'' points resulting in the antiferromagnetic order which is suppressed due to the frustration effect in the case without the Na order. 
% Our results are consistent with the experimental result that the resistivity shows no anomaly at $T_{c1}$, while rapidly increase below $T_{c2}$ \cite{Foo}. 

In the present work, the antiferromagnetism and the orbital order according to the charge order due to the effect of $V$ are strengthened each other, and the coexistence state is obtained. 
However, there is a possibility that the coexistence state is obtained 
only due to the effect of $U$ without $V$, where the cooperation between the antiferromagnetic order and the orbital order is a crucial role. 
Explicit results for such state and detailed comparisons with experimental results will be reported in a subsequent paper. 
% We propose the possibility that the origin of the metal-insulator transition at $T_{c2}$ is a transition to the coexistence state between antiferromagnetic, charge and orbital ordered state. 
% Detailed comparisons with experimental results and a calculation of $T$ dependence are further tasks. 

%%%%%%%%%%%%%%%%%%%%%%%%%%%%%%%%%%%%%%%%%%%%%%%%%%%%%%%%%%%%%%%%%%%%%%%%%%%%
\section*{Acknowledgements}
The authors thank M. Sato, Y. Kobayashi, M. Yokoi and T. Moyoshi for many useful comments and discussions.
This work was performed under the interuniversity cooperative Research program of the Institute for Materials Research, Tohoku University, and was supported by the Grant-in-Aid for Scientific Research from the Ministry of Education, Culture,  Sports, Science  and Technology. 

%%%%%%%%%%%%%%%%%%%%%%%%%%%%%%%%%%%%%%%%%%%%%%%%%%%%%%%%%%%%%%%%%%%%%%%%%%%%

\end{document}